\documentclass[sigconf,10pt]{acmart}

\usepackage[normalem]{ulem}
\usepackage{epstopdf}
\usepackage{tikz}
\usepackage{pifont}
\usepackage{endnotes}
\usepackage{amsmath,amsfonts}
\usepackage{graphicx}
\usepackage{textcomp}
\usepackage{multirow}
\usepackage{booktabs}
\usepackage{hyphenat}
\usepackage{fontawesome}
\usepackage{color,xcolor}
\usepackage{algpseudocode}
\usepackage{algorithm}
\usepackage{subfigure}
\usepackage{array}
\usepackage{enumerate,enumitem}
\usepackage{makecell}
\usepackage{balance}
\usepackage{filecontents}
\usepackage{xspace}
\usepackage{caption}
\usepackage{fancyhdr}
\usepackage{setspace}
\usepackage{listings}
\usepackage{environ}
\usepackage[]{hyperref}

\newcommand{\sys}{\lowercase{d}-HNSW\xspace}

\renewcommand\footnotetextcopyrightpermission[1]{}

\setcopyright{none}

\begin{document}

\title{Efficient Vector Search on Disaggregated Memory with \sys}

\author{Yi Liu*}
\affiliation{
  \institution{University of California Santa Cruz}
  \city{Santa Cruz}
  \state{CA}
  \country{USA} 
}
\author{Fei Fang*}
\affiliation{
  \institution{University of California Santa Cruz}
  \city{Santa Cruz}
  \state{CA}
  \country{USA}
}
\author{Chen Qian}
\affiliation{
  \institution{University of California Santa Cruz}
  \city{Santa Cruz}
  \state{CA}
  \country{USA}
}
\thanks{* Equal contribution.}


\renewcommand{\shortauthors}{Yi Liu, Fei Fang, Chen Qian\\
University of California, Santa Cruz}
\begin{abstract}
Efficient vector query processing is essential for powering large-scale AI applications, such as LLMs. However, existing solutions struggle with growing vector datasets that exceed the memory capacity of a single machine, leading to excessive data movement and resource underutilization in monolithic architectures.

We introduce \sys, the first vector search engine for RDMA-based disaggregated memory systems. \sys achieves high performance by supporting efficient data indexing with minimal network communication overhead.
At its core, \sys introduces a novel disaggregation of the HNSW graph-based vector index, leveraging the properties of greedy search to coordinate data transfers efficiently between the memory and compute pools. Specifically, \sys incorporates three key techniques:
(i) Representative index caching, which constructs a lightweight index from a sampled subset of the data and caches it in the compute pool to minimize frequent access to critical components of the hierarchical graph index;
(ii) RDMA-friendly data layout, which optimizes data placement to reduce networking round trips for both vector queries and insertions; and
(iii) Batched query-aware data loading, which mitigates bandwidth usage between memory and compute pools, addressing the limited cache capacity of compute nodes.
The experimental results demonstrate that \sys outperforms Naive \sys implementation by up to 117$\times$ in query latency while maintaining a recall of 0.87 on the SIFT1M dataset.
\end{abstract}


\maketitle
\pagestyle{plain}

\vspace{-2ex}
\section{Introduction}
\label{sec:intro}

Vector similarity search~\cite{wang2021milvus,wang2024vector,hm-ann,zhang2019grip,rummy,auncel} aims to identify the most similar vectors from a large dataset given a query vector. Approximate nearest neighbor (ANN) search has emerged as a practical alternative for exact top-$k$ nearest neighbor search, offering significant speedups with minimal accuracy trade-offs. Vector databases leverage ANN search techniques to support efficient retrieval, making them essential for a wide range of ML applications~\cite{faiss,liu2024retrievalattention}, including search engines~\cite{pan2024survey} and recommendation systems\cite{wentao}. 
In particular, they play a critical role in powering large language models (LLMs) and retrieval-augmented generation (RAG) systems~\cite{ragcache} by enabling fast, high-dimensional similarity searches over massive embedding spaces. In RAG, a vector database retrieves semantically relevant documents based on the user prompt’s embedding, allowing LLMs to generate responses with external knowledge rather than limited to the information encoded in their model parameters. As these AI-driven applications~\cite{wentao, memserve} continue to grow, the demand for scalable and high-performance vector search solutions has become increasingly crucial.

\begin{figure}[!t]
	\centering
	\includegraphics[scale=0.66]{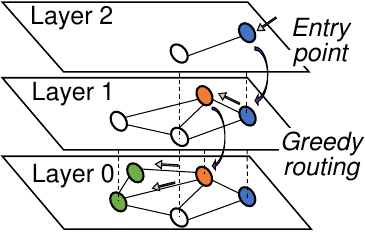}
	\caption{Graph-based vector search index: HNSW.}
	\vspace{-3ex}
	\label{fig:hnsw}
\end{figure}

Disaggregation~\cite{wang2023disaggregated} is gaining attention in cloud computing by separating storage and compute hardware resources, allowing them to scale independently for better flexibility and efficiency. Different resource pools are connected with high-speed networks to realize fast data transfer. For example, RDMA (Remote Direct Memory Access)~\cite{guideline} is one of the high-performance fabrics that enables direct memory access between remote machines, bypassing the CPU to reduce latency and improve throughput.
Recent industry efforts, such as DeepSeek's 3FS~\cite{3fs,deepseek}, have demonstrated the benefits of leveraging RDMA for AI training and inference, enabling high-speed remote memory access with low latency. Inspired by this architectural shift~\cite{tsai2020disaggregating,wang2023disaggregated}, we are motivated to propose an RDMA-based disaggregated vector database designed to improve hardware resource utilization with the high-throughput vector query.

Among the various similarity search algorithms~\cite{kdtree,lsh,wang2024vector}, graph-based approaches~\cite{nsg,hnsw} have demonstrated superior performance in both recall and latency. Hierarchical Navigable Small World (HNSW)~\cite{hnsw} is a widely adopted graph-based index that balances vector search accuracy and efficiency. 
Thus, in this work, we introduce \sys, a fast, RDMA-based vector similarity search engine designed for disaggregated memory system. \sys aims to bridge the gap between high-performance vector similarity search and the emerging disaggregated architecture in datacenter, ensuring scalability and efficiency in handling high-throughput data queries. 

The disaggregated memory pool provides abundant memory resources, allowing us to store both the HNSW index and all original floating-point vector data on it. Intuitively, the disaggregated compute instances handle data requests and reply on one-sided RDMA primitives to directly access the index and vectors, bypassing memory instances' CPUs. However, this approach presents several challenges.
(i) The greedy algorithm~\cite{lighttraffic} in HNSW navigates the index by comparing distances to the query vector along a search path, where each node on the graph represents a vector. The traversal path is unpredictable, and if we need to read vectors on each single step along the path via the network, the number of round trips required for a vector query becomes excessive. To mitigate this, we propose partitioning vectors into groups with an \textit{representative index} and selectively reading only the partitions that most likely contain top-$k$ candidates. 
(ii) All partitions are compactly serialized and written to remote registered memory. When a new vector is inserted, it needs to be stored in available memory while ensuring fast index access for different partitions. If we allocate a global memory space for inserted vectors, those belonging to the same partition will be scattered across fragmented memory regions, leading to high latency as the RDMA NIC needs to issue multiple network or PCIe round trips. To address this, we propose an \textit{RDMA-friendly graph index layout} that enables efficient data queries while supporting dynamic vector insertions. 
(iii) Since we process vector queries in a batch~\cite{faiss} and there is limited cache DRAM space in the compute pool, we propose \textit{query-aware data loading} to reduce partition data loading from the memory pool and save bandwidth by pruning duplicate partition transfers, thereby improving vector query throughput.

We implement a prototype of \sys with 12K LoC and evaluate it against other RDMA-based approaches in terms of vector query recall, latency, and throughput across various datasets. The results show \sys outperforms other baselines by up to 117$\times$ with top-10 benchmarking. 
\textbf{To our best knowledge, \sys is the first vector database designed for RDMA-based disaggregated memory systems. We believe \sys can inspire researchers to explore this area further and propose new solutions to boost the performance of disaggregated vector databases.}

\vspace{-2ex}
\section{Background}
\label{sec:background}
\vspace{-.5ex}

\subsection{Graph-based vector search with HNSW.}
\label{sec:background:vectordb}

Vector similarity search is crucial for efficiently retrieving high-dimensional data in modern ML applications such as RAG~\cite{ragcache} for LLMs. 
Traditional methods like KD-trees~\cite{kdtree} and LSH~\cite{lsh} struggle with scalability and search accuracy in high-dimensional spaces, leading to the development of graph-based indexing techniques~\cite{hnsw,nsg}. 
These methods construct a navigable graph where data points serve as nodes, and edges encode proximity relationships, enabling fast traversal during queries. 
For example, as shown in Fig.~\ref{fig:hnsw}, HNSW builds a multi-layered graph~\cite{hnsw} where upper layers provide a coarse-grained overview for fast entry into the structure, and lower layers refine the search with more densely connected nodes. 
During a query, the search starts from an \textit{entry point} and follows a greedy routing strategy, moving to the closest neighbor at each layer. This closest vector will become the entry point to the next layer, performing greedy routing again toward the queried vector while refining the candidate set.
The number of vectors in each layer increases \textit{exponentially}.
By leveraging small-world properties and efficient greedy search heuristics, HNSW significantly improves both recall and query speed compared to earlier graph-based methods, making it one of the most effective ANN search algorithms in modern vector databases.

\vspace{-1.ex}
\subsection{RDMA-based disaggregated memory.}
\label{sec:background:rdma}
RDMA technologies (e.g. RoCE~\cite{roce}, Infiniband~\cite{guideline}) enable reliable and in-order packet delivery, making them well-suited for indexing structures in disaggregated memory systems. 
It supports RDMA READ/WRITE for fetching and writing data directly on remote memory without CPU involvement, and atomic operations like Compare-And-Swap (CAS) and Fetch-And-Add (FAA), enable efficient and lock-free data access.
Designing an efficient indexing data structure tailored for RDMA-based remote memory applications can reduce system computation overheads, minimize network round trips, and realize data access with low latency.

\vspace{-1.ex}
\section{\sys Design}
\label{sec:design}

\begin{figure}[!t]
	\centering
	\includegraphics[scale=0.41]{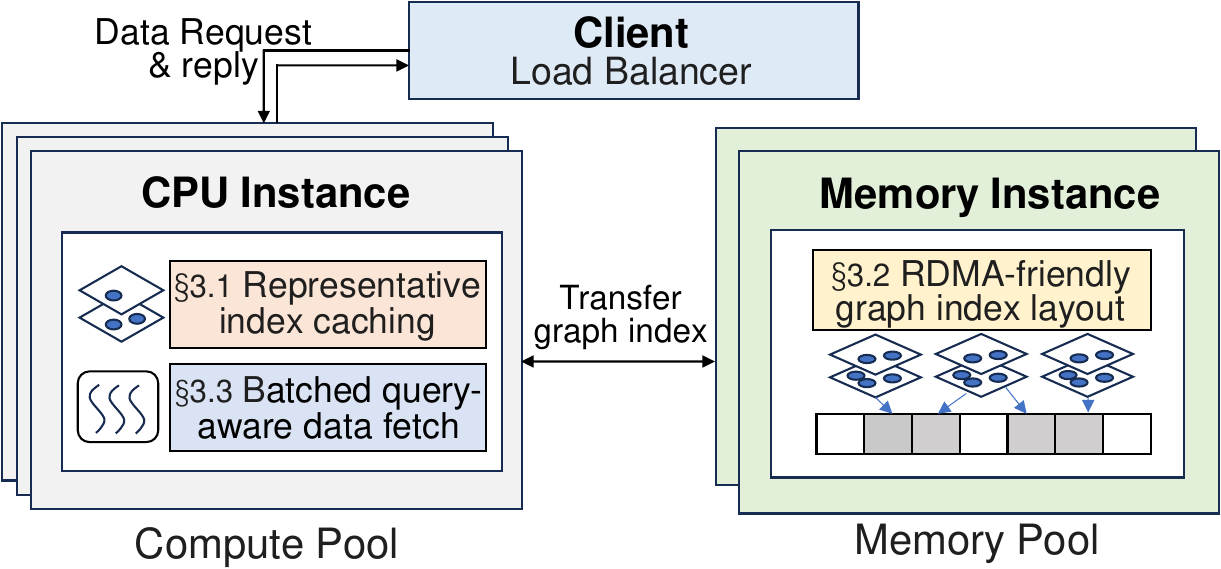}
	\caption{The overview of \sys.}
	\vspace{-3ex}
	\label{fig:overview}
\end{figure}

We present \sys, an RDMA-based vector similarity search engine on disaggregated memory. 
\sys exploits the characteristics of RDMA-based memory data accessing and graph-based index HNSW to realize fast and bandwidth-efficient vector query processing. 
\sys achieves so by representative index caching (\S\ref{sec:design:cache}), RDMA-friendly graph index storage in remote memory (\S\ref{sec:design:layout}), and query-aware batched data loading (\S\ref{sec:design:load}).
Here, we provide a brief overview of \sys as Fig.~\ref{fig:overview} shows, \sys requires tailored coordination between compute instances and memory instances on vector query serving. 
We assume the client load balancer distributes the workload across multiple CPU instances. The compute and memory pools are interconnected via RDMA, enabling efficient transfer of vector indices and data.
\textbf{We target the disaggregated scenario where compute pools contain abundant CPU resources across many instances, each with limited DRAM serving as a cache, while memory instances have extremely weak computational power, handling lightweight memory registration tasks.}

\subsection{Representative index caching.}
\label{sec:design:cache}

\begin{figure}[!t]
	\centering
	\includegraphics[scale=0.345]{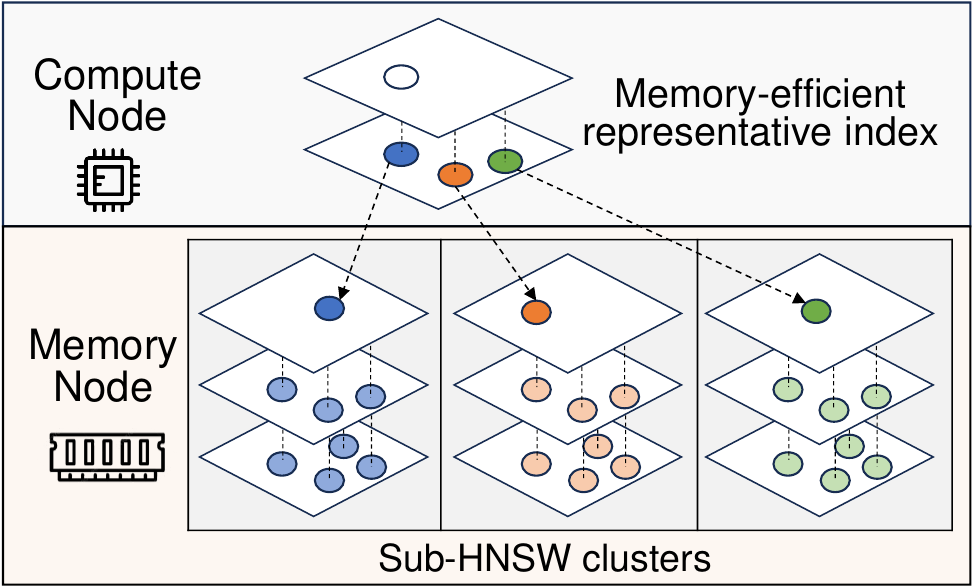}
	\caption{Representative index caching in \sys.}
	\vspace{-2ex}
	\label{fig:cache}
\end{figure}
Graph-based vector search schemes~\cite{nsg, hnsw} rely on greedy routing to iteratively navigate toward the queried vector. However, the search path can span the entire graph, potentially covering distant vectors. For example, HNSW exhibits small-world properties, allowing long-range connections between vectors that are far apart. However, loading the entire graph index from the memory pool to the computer pool for each query is impractical, because the compute pool has limited storage resources in a disaggregated system. This approach would not only consume excessive bandwidth by transferring a significant portion of untraversed vectors but also introduce additional latency, thereby degrading the overall search efficiency.

We propose partitioning the vector database into multiple subsets, as shown in Fig.~\ref{fig:cache}. Inspired by Pyramid~\cite{pyramid}, we construct a three-layer representative HNSW, referred to as \textit{meta-HNSW}, by uniformly selecting 500 vectors. This meta-HNSW serves as a lightweight index and a cluster classifier for the entire dataset, and it only costs 0.373 MB for SIFT1M and 1.960 MB for GIST1M datasets from our experiments.
The search process starts from a fixed entry point in the top layer $L_2$ of meta-HNSW and applies greedy routing at each layer, traversing downward until reaching a vector in its bottom layer $L_0$. Each vector in $L_0$ defines a partition and serves as an entry point to a corresponding \textit{sub-HNSW}. All vectors assigned to the same partition will be used to construct their respective sub-HNSW.
The overall graph index consists of two components: meta-HNSW, which provides coarse-grained classification, and sub-HNSWs, which enable fine-grained search within partitions. 
\textbf{To improve search efficiency in disaggregation, we cache the lightweight meta-HNSW in the compute pool, allowing it to identify the most relevant sub-HNSW clusters for a given query.} 
Meanwhile, we put all sub-HNSW clusters in the memory pool. For each vector query, only a small subset of sub-HNSW clusters needs to be loaded from the memory pool via the network, reducing both bandwidth usage and search latency.

\subsection{RDMA-friendly graph index storage layout in remote memory.}
\label{sec:design:layout}

\begin{figure}[!t]
	\centering
	\includegraphics[scale=0.44]{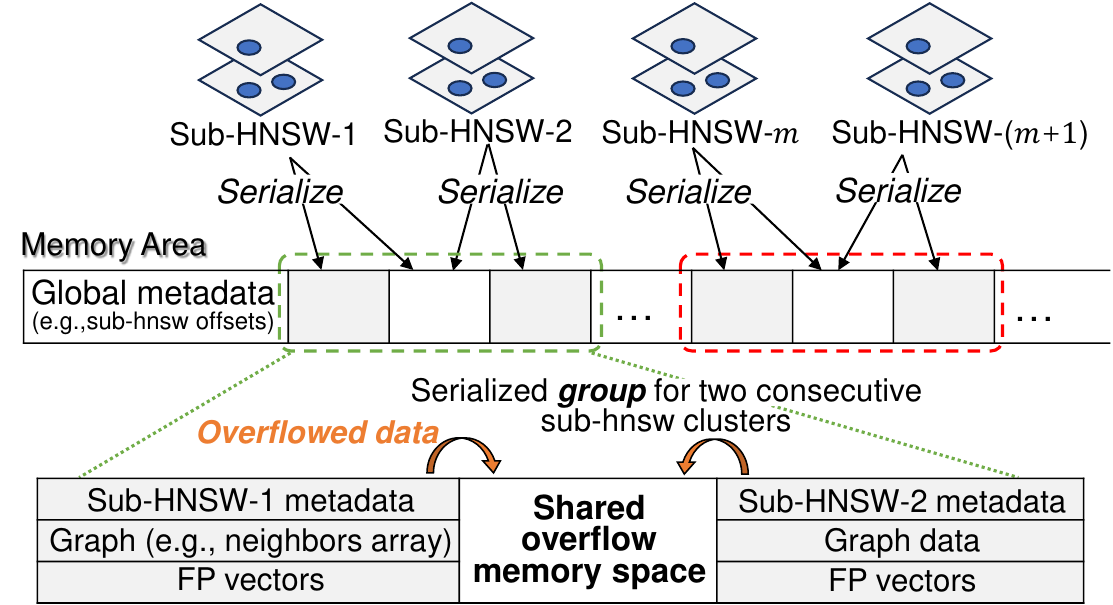}
	\caption{RDMA-friendly sub-HNSW indexing data layout in remote memory.}
	\vspace{-2ex}
	\label{fig:layout}
\end{figure}

RDMA enables efficient data access to targeted remote memory addresses. To efficiently read and write sub-HNSW cluster data in remote memory, an intuitive approach is to serialize all sub-HNSW clusters in the registered memory. 
Given that the top-$m$ closest sub-HNSW clusters for a queried vector $q$ are \{$S_0,..,S_{m-1}$\}, the compute instance can issue RDMA\_READ commands to access these serialized clusters and then deserialize them. 
However, two challenges arise: (1) If the queried clusters \{$S_0,..,S_{m-1}$\} are not stored contiguously in memory, multiple RDMA round trips are required, increasing latency. (2) When new vectors are inserted, the size of each sub-HNSW cluster may exceed the allocated space. Since shifting all stacked sub-HNSW clusters is impractical, newly inserted vectors and their metadata may be placed in non-contiguous memory regions if they are simply appended at the tail of the available area. This fragmentation increases access latency and reduces query throughput due to the higher cost of scattered index access. 

As shown in Fig.~\ref{fig:layout}, we allocate and register a continuous memory space in memory instance to store both the serialized HNSW index and floating-point vectors. At the beginning of this memory space, a global metadata block records the offsets of each sub-HNSW cluster, as their sizes vary. The remaining memory space is divided into \textit{groups}, each of which is capable of holding two sub-HNSW clusters.
Within each group, the first section stores the first serialized sub-HNSW cluster, which includes its metadata, neighbor array for HNSW, and the associated floating-point vectors. The second sub-HNSW cluster is placed at the end of the group. Between these two clusters, we allocate a 0.75 MB for SIFT1M 3.92 MB for GIST1M shared overflow memory space to accommodate newly inserted vectors for both sub-HNSW clusters.
When a vector query requires loading a sub-HNSW cluster, the compute instance issues an RDMA\_READ command to retrieve the cluster along with its corresponding shared overflow memory space. This layout ensures that newly inserted vectors are stored continuously with the original sub-HNSW data, enabling them to be read back with a one-time RDMA\_READ command. To optimize memory usage, each pair of adjacent sub-HNSW clusters shares a single overflow memory space for accommodating newly inserted vectors rather than allocating a separate one for each cluster.

If multiple sub-HNSW clusters need to be loaded into the compute pool for batched query processing, and they are not stored continuously in memory, we leverage \textit{doorbell batching} to read them in a single network round-trip with RDMA NIC issuing multiple PCIe transactions. However, there is a tradeoff in the number of batched operations within a single RDMA command. If too many operations are included in one round-trip, it can interfere with other RDMA commands and incur long latency due to the scalability of the RDMA NIC. 
The memory offsets of each sub-HNSW cluster are cached in all compute instances after the sub-HNSW clusters are written to the memory pool, with the latest version stored at the beginning of the memory space in the memory instance.

\vspace{-1ex}
\subsection{Query-aware batched data loading.}
\label{sec:design:load}

\begin{figure}[!t]
	\centering
	\includegraphics[scale=0.36]{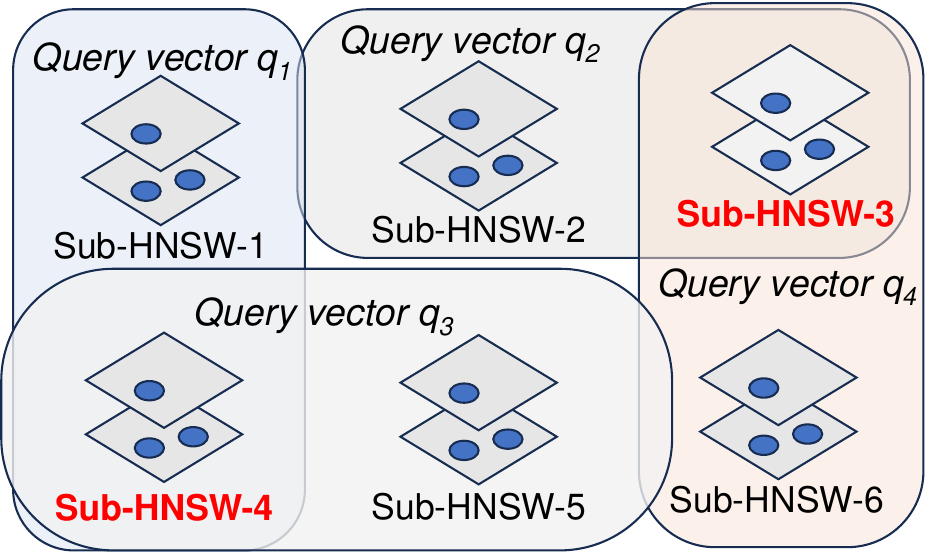}
    \vspace{-1ex}
	\caption{Query-aware sub-HNSW clusters loading.}
	\vspace{-2ex}
	\label{fig:load}
\end{figure}
To reduce bandwidth usage for transferring graph index and improve query efficiency, we propose merging sub-HNSW index loading for queried vectors in the same batch.

Given a batch of queried vectors \{$q_1$,$q_2$, ...,$q_s$\} and a total of $m$ sub-HNSW clusters, each queried vector requires searching the top-$k$ closest vectors from the $b$ closest sub-HNSWs. However, the DRAM resources in the compute instance can only accommodate and cache $c$ sub-HNSWs. To optimize loading, we analyze the required $b*s$ sub-HNSWs \textit{online} and ensure that each sub-HNSW is loaded from the memory pool only \textbf{once}.

For example, as shown in Fig.~\ref{fig:load}, queried vector $q_1$’s two closest sub-HNSW clusters are $S_1$ and $S_4$, while $q_3$’s two closest sub-HNSWs are $S_4$ and $S_5$. Similarly, $S_3$ is required for both $q_2$ and $q_4$. Given a doorbell batch size of 2 for accessing sub-HNSWs, the compute instance can issue an RDMA\_READ command to fetch index $S_3$ and $S_4$ in one network round-trip, then compute the top-$k$ closest vectors candidates for all queries \{$q_1$,$q_2$,$q_3$,$q_4$\} first. The results will be temporarily stored for further computation and comparison because each query vector still requires another sub-HNSW to obtain the final answer. Note that $S_3$ and $S_4$ will not be loaded again within the same batch.

Once all required sub-HNSW clusters for the batched queried vectors have been loaded and traversed, the query results will be returned. Additionally, we retain the most recently loaded $c$ sub-HNSWs for the next batch. If the required sub-HNSWs are already in the compute instance, they do not need to be loaded again, further reducing data transfer overhead.

\vspace{-3.ex}
\section{Evaluation}
\label{sec:eval}

\begin{figure*}[!t]
\centering
\renewcommand\thesubfigure{}
\subfigure[\hspace{4mm}(a) SIFT1M@10.]{
    \label{fig:eval:search:a}
    \includegraphics[width=0.252\textwidth]{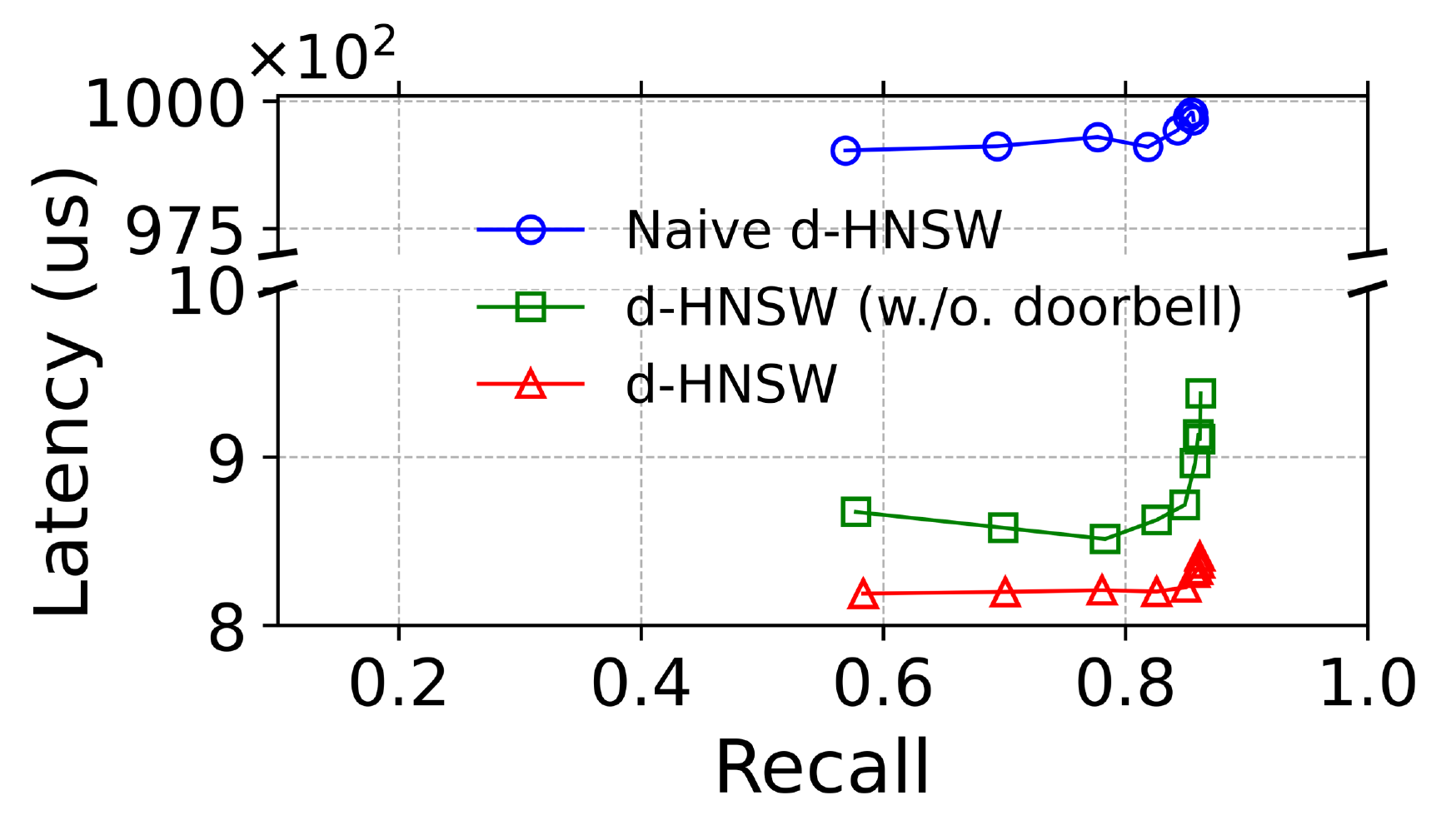}}
\hspace{-2.4ex}
\subfigure[\hspace{4mm}(b) SIFT1M@1.]{
    \label{fig:eval:search:b}
    \includegraphics[width=0.252\textwidth]{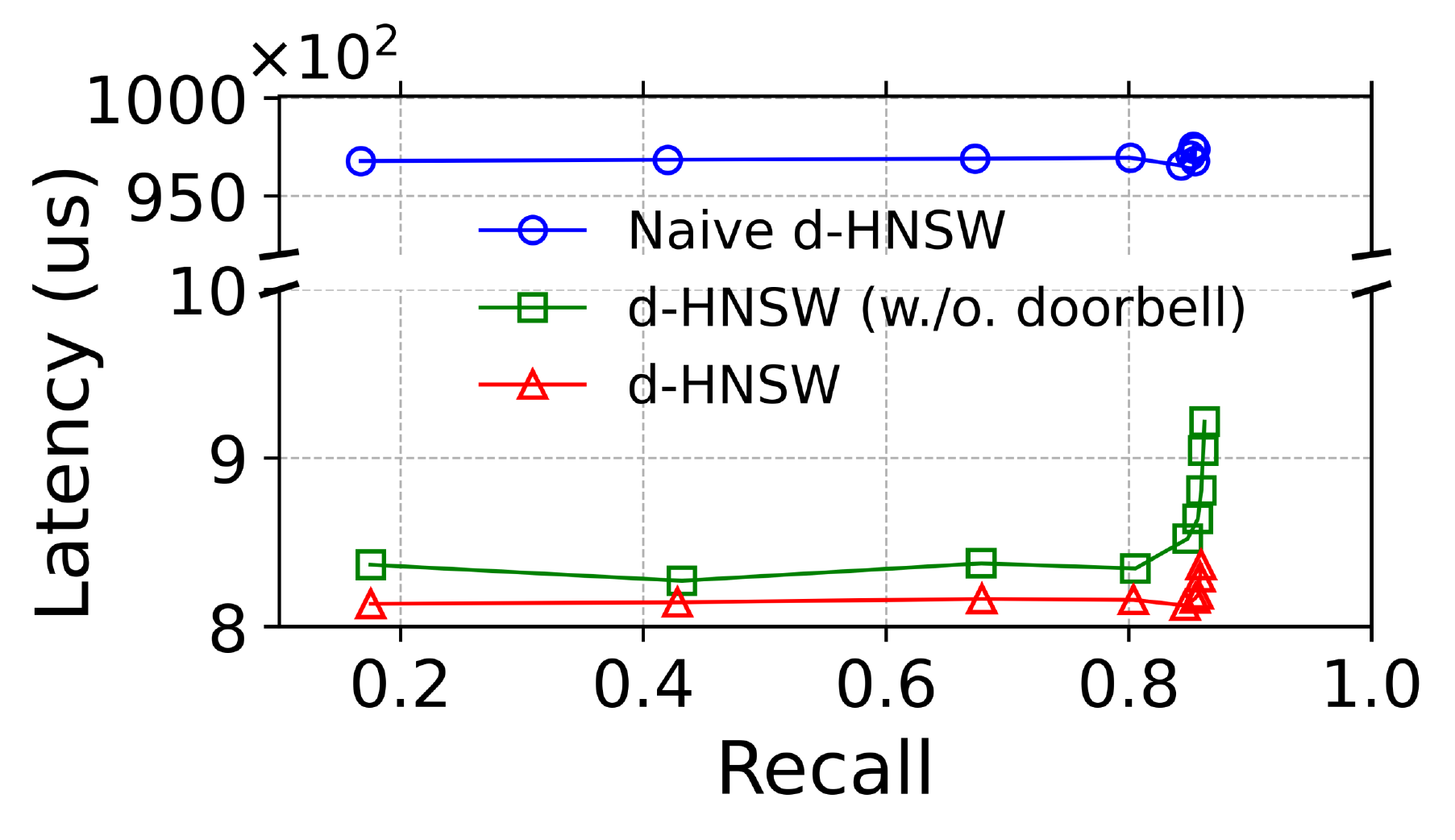}}
\hspace{-2.4ex}
\subfigure[\hspace{4mm}(c) GIST1M@10.]{
    \label{fig:eval:search:c}
    \includegraphics[width=0.252\textwidth]{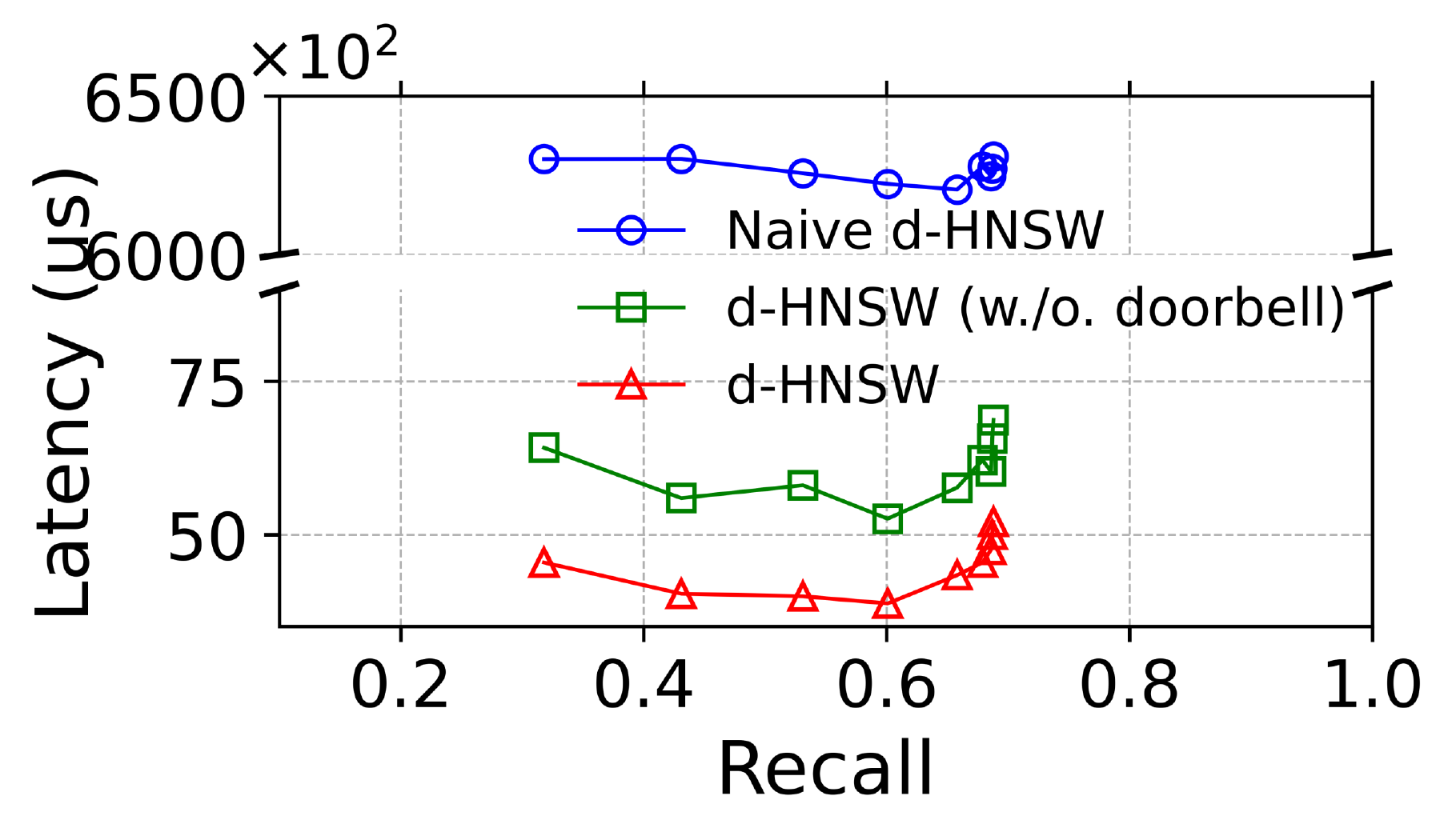}}
\hspace{-2.4ex}
\subfigure[\hspace{4mm}(d) GIST1M@1.]{
    \label{fig:eval:search:d}
    \includegraphics[width=0.252\textwidth]{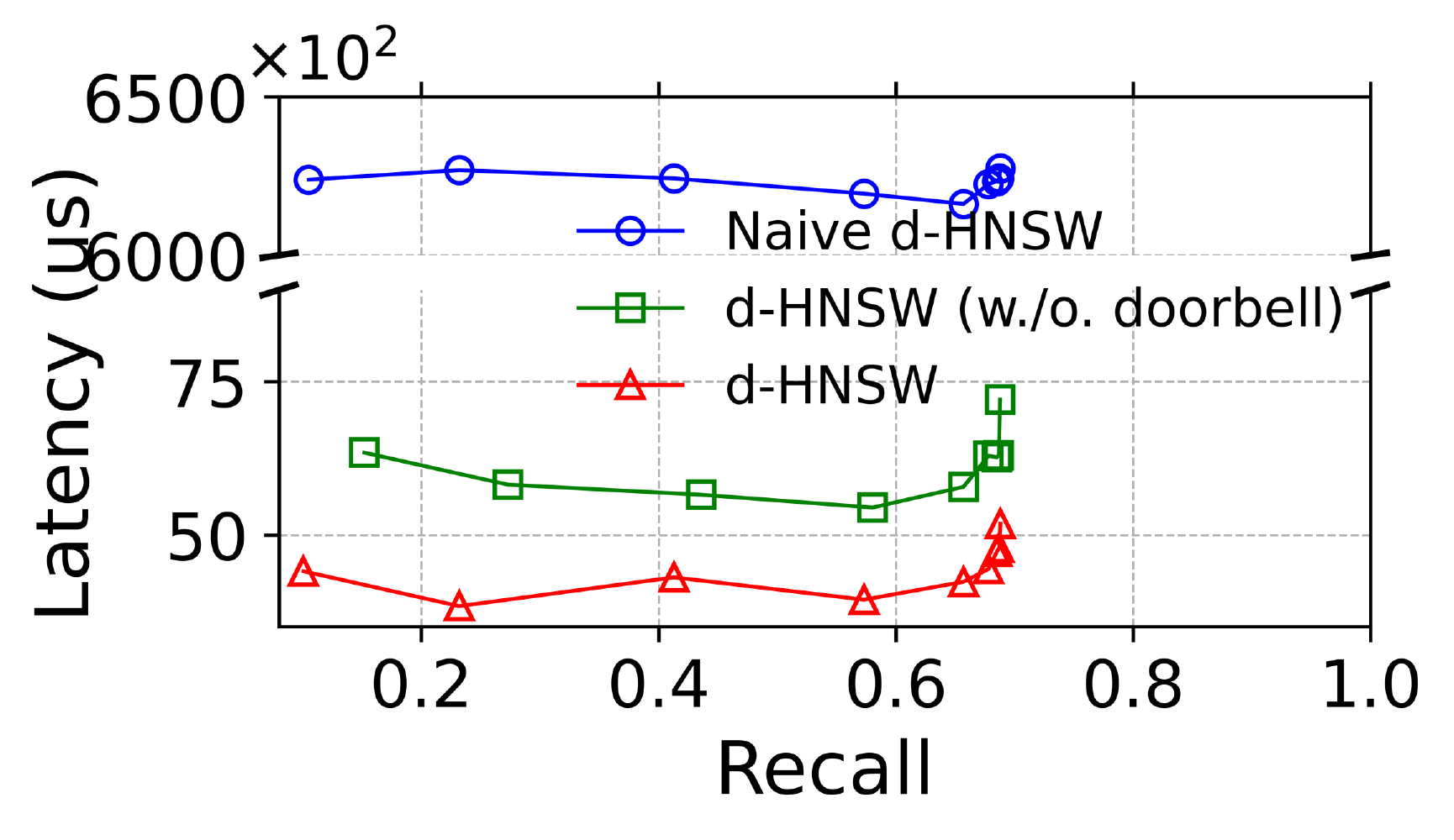}}
\vspace{-3ex}
\caption{Latency-recall evaluation of \sys and baselines.}
\label{fig:eval:search}
\vspace{-3ex}
\end{figure*}

We develope and evaluate the prototype of \sys on CloudLab~\cite{cloudlab} using real-world hardware. Our testbed consists of four Dell PowerEdge R650 servers, each equipped with two 36-core Intel Xeon Platinum CPUs, 256GB RAM, a 1.6TB NVMe SSD, and a Mellanox ConnectX-6 100Gb NIC. 
Three servers act as a compute pool, while one serves as the memory instance. We compare \sys against the following baselines:
(1) Native-HNSW: when a vector query arrives at a compute node, Native-HNSW issues an RDMA read command to fetch corresponding sub-HNSW clusters, bypassing the memory node's CPU.
(2) \sys (w./o. doorbell): with meta-HNSW caching and query-aware data loading, the compute node reads sub-HNSW clusters in multiple round-trips, while our \sys reads discontinuous sub-HNSW clusters in a single doorbell batch.

Each server has 144 hyperthreads, which are divided into 8 compute instances. Each instance runs a vector query worker that sends RDMA commands for top-k vector retrieval. Each instance uses 18 threads for OpenMP parallel HNSW search. The cache in each compute instance is configured to store only 10\% of the total sub-HNSW clusters in the memory pool. At runtime, the batch size for vector queries is set to 2000. 

\textbf{Latency-recall curve evaluation.}
We evaluate \sys and the baselines using the SIFT1M and GIST1M datasets, setting the top-$k$ parameter as top-1 and top-10, respectively. All compute instances across three servers issue vector queries to the memory instance together. 

Fig.~\ref{fig:eval:search} presents the latency-recall curves for all three schemes with the varied $efSearch$ from 1 to 48. The $efSearch$ parameter determines the number of dynamic candidates maintained during the sub-HNSW search process. 
In the SIFT1M dataset with top-10 vector query shown in Fig.~\ref{fig:eval:search:a}(a), \sys reduces latency by up to 117$\times$ and 1.12$\times$ compared to naive \sys and \sys without doorbell, respectively, while achieving a recall of approximately 0.86 when $efSearch$ reaches 48. The reason is naive \sys issues an RDMA read round-trip to access each involved sub-HNSW cluster. Also, the doorbell mechanism further optimizes performance by batching memory accesses across multiple fragmented addresses within a single round-trip.
For top-1 query in SIFT1M shown in Fig.~\ref{fig:eval:search:b}(b), the upper recall reaches 0.85 when $efSearch$ is set to 48. The latency is further reduced across all schemes since only the closest vector needs to be selected.

Similarly, in the GIST1M dataset shown in Figs.~\ref{fig:eval:search:c}(c)(d), \sys achieves up to 121$\times$ and 1.30$\times$ lower latency compared to naive \sys and \sys without doorbell, respectively. Due to the higher dimensionality of GIST1M vectors, query latency is generally higher than in SIFT1M.

\textbf{Latency breakdown of vector query.}
We break down the latency of each scheme to analyze the source of \sys's performance advantage. The total latency of a vector query consists of three components: data transfer over the network, meta-HNSW (cache) computation, and sub-HNSW computation on loaded data.
Table~\ref{tab:eval:sift} presents the latency breakdown for the SIFT1M dataset with top-1 queries. \sys benefits from significantly reduced network latency, measured at 527$\mu$s, which is 0.005$\times$ and 0.84$\times$ lower than that of naive \sys and \sys without doorbell, respectively. 
The number of round-trips per vector query are 3.547 for naive d-HNSW and 0.896 for d-HNSW w./o. doorbell, $4.75 \times 10^{-3}$ for d-HNSW for SIFT1M.
Similarly, as shown in Table.~\ref{tab:eval:gist}, \sys also achieves the lowest network latency in the dataset GIST1M.

\begin{table}[t]
    \centering
    \renewcommand{\arraystretch}{1.1}
    \setlength{\tabcolsep}{2pt}
    \fontsize{9}{9}\selectfont
    \begin{tabular}{cccc}
        \toprule
        \textbf{Scheme}  &  Network & Sub-HNSW & Meta-HNSW\\
        \midrule
        Naive \sys & 90271.2$\mu$s & 6564.5$\mu$s & 13.52$\mu$s \\
        \sys (w./o. doorbell) & 607.5$\mu$s & 287.0$\mu$s & 9.97$\mu$s \\
        \sys & \textbf{\underline{527.6$\mu$s}} & 269.2$\mu$s & 9.75$\mu$s \\
        \bottomrule
    \end{tabular}
    \vspace{1ex}
    \caption{Latency breakdown for SIFT1M@1 with efSearch as 48.}
    \label{tab:eval:sift}
    \vspace{-8.5ex}
\end{table}

\begin{table}[t]
    \centering
    \renewcommand{\arraystretch}{1.1}
    \setlength{\tabcolsep}{2pt}
    \fontsize{9}{9}\selectfont
    \begin{tabular}{cccc}
        \toprule
        \textbf{Scheme}  &  Network & Sub-HNSW & Meta-HNSW\\
        \midrule
        Naive \sys & 422.9ms & 35.3ms & 61.1$\mu$s \\
        \sys (w./o. doorbell) & 2.9ms & 1.27ms & 52.6$\mu$s \\
        \sys & \textbf{\underline{1.3ms}} & 1.48ms &  46.9$\mu$s \\
        \bottomrule
    \end{tabular}
    \vspace{1ex}
    \caption{Latency breakdown for GIST1M@1 with $efSearch$ as 48.}
    \label{tab:eval:gist}
    \vspace{-2ex}
\end{table}

\vspace{-1ex}
\section{Related Work}
\vspace{-1ex}
\label{sec:relatedwork}

\textbf{Disaggregated memory system.}
Disaggregated memory systems have recently received essential attention due to enabling flexible resource allocation and improving hardware utilization in data centers. Existing work studies various solutions to managing and developing memory disaggregation from various systematic views, including architectural support~\cite{shen2023fusee,tsai2020disaggregating,wang2023disaggregated,wang2022case}, operating systems~\cite{shan2018legoos,wang2020semeru}, KVCache management for LLMs~\cite{memserve,mooncake}, disaggregated KV stores~\cite{zuo2022race,li2023rolex,luo2023smart,lu2024dex,luo2024chime,shen2023ditto,wang2022sherman,outback}, transactional systems~\cite{zhang2022ford, zhang2024motor}, in-network computation systems~\cite{wang2024rcmp, cowbird, hidpu, wei2023characterizing}. 
\sys is orthogonal to these works. 

\noindent\textbf{Approximate similarity search system.}
Approximate similarity search has become a fundamental technique for efficiently retrieving high-dimensional data vectors. 
Various algorithms have been developed to balance search efficiency and accuracy, like KD-trees~\cite{kdtree}, graph-based search structures~\cite{hnsw}, and quantization techniques~\cite{faiss}. 
Furthermore, advancements in hardware acceleration, such as GPU-based indexing~\cite{rummy} and CXL-based indexing~\cite{jang2023cxl}, have further improved search performance. As data volumes continue to grow, optimizing vector search for both accuracy and resource efficiency remains an active research area. This includes adaptive search strategies~\cite{li2020improving, zhang2019grip} and storage tier-aware optimizations~\cite{hm-ann} to meet different service level objectives (SLOs)~\cite{auncel,su2024vexless}.

\vspace{-2ex}
\section{Conclusion}
\label{sec:conclusion}
\vspace{-.5ex}
We present \sys, the first RDMA-based vector similarity search engine designed for disaggregated memory. 
\sys enhances vector request throughput and minimizes data transfer overhead by implementing an RDMA-friendly data layout for memory nodes. 
Additionally, \sys optimizes batched vector queries by eliminating redundant vector transfers for batched vector queries.
We evaluate \sys on various datasets and show its latency outperforms other RDMA-based baselines by up to 117$\times$ with top-10 benchmarking.

\bibliographystyle{plain}
\bibliography{sample-base}

\begin{thebibliography}{10}

\bibitem{cowbird}
Xinyi Chen, Liangcheng Yu, Vincent Liu, and Qizhen Zhang.
\newblock Cowbird: Freeing cpus to compute by offloading the disaggregation of
  memory.
\newblock In {\em Proceedings of the ACM SIGCOMM 2023 Conference}, pages
  1060--1073, 2023.

\bibitem{cloudlab}
scientific infrastructure for research on the future of cloud~computing.
  CloudLab:~Flexible.
\newblock https://www.cloudlab.us.

\bibitem{deepseek}
DeepSeek.
\newblock https://www.deepseek.com/.

\bibitem{pyramid}
Shiyuan Deng, Xiao Yan, KW~Ng Kelvin, Chenyu Jiang, and James Cheng.
\newblock Pyramid: A general framework for distributed similarity search on
  large-scale datasets.
\newblock In {\em 2019 IEEE International Conference on Big Data (Big Data)},
  pages 1066--1071. IEEE, 2019.

\bibitem{3fs}
Fire flyer~file system.
\newblock https://github.com/deepseek-ai/3fs/.

\bibitem{nsg}
Cong Fu, Chao Xiang, Changxu Wang, and Deng Cai.
\newblock Fast approximate nearest neighbor search with the navigating
  spreading-out graphs.
\newblock {\em {PVLDB}}, 12(5):461 -- 474, 2019.

\bibitem{lsh}
Aristides Gionis, Piotr Indyk, Rajeev Motwani, et~al.
\newblock Similarity search in high dimensions via hashing.
\newblock In {\em Vldb}, volume~99, pages 518--529, 1999.

\bibitem{memserve}
Cunchen Hu, Heyang Huang, Junhao Hu, Jiang Xu, Xusheng Chen, Tao Xie, Chenxi
  Wang, Sa~Wang, Yungang Bao, Ninghui Sun, et~al.
\newblock Memserve: Context caching for disaggregated llm serving with elastic
  memory pool.
\newblock {\em arXiv preprint arXiv:2406.17565}, 2024.

\bibitem{jang2023cxl}
Junhyeok Jang, Hanjin Choi, Hanyeoreum Bae, Seungjun Lee, Miryeong Kwon, and
  Myoungsoo Jung.
\newblock Cxl-anns:software-hardware collaborative memory disaggregation and
  computation for billion-scale approximate nearest neighbor search.
\newblock In {\em 2023 USENIX Annual Technical Conference (USENIX ATC 23)},
  pages 585--600, 2023.

\bibitem{ragcache}
Chao Jin, Zili Zhang, Xuanlin Jiang, Fangyue Liu, Xin Liu, Xuanzhe Liu, and Xin
  Jin.
\newblock Ragcache: Efficient knowledge caching for retrieval-augmented
  generation.
\newblock {\em arXiv preprint arXiv:2404.12457}, 2024.

\bibitem{guideline}
Anuj Kalia, Michael Kaminsky, and David~G Andersen.
\newblock Design guidelines for high performance rdma systems.
\newblock In {\em 2016 USENIX annual technical conference (USENIX ATC 16)},
  pages 437--450, 2016.

\bibitem{li2020improving}
Conglong Li, Minjia Zhang, David~G Andersen, and Yuxiong He.
\newblock Improving approximate nearest neighbor search through learned
  adaptive early termination.
\newblock In {\em Proceedings of the 2020 ACM SIGMOD International Conference
  on Management of Data}, pages 2539--2554, 2020.

\bibitem{li2023rolex}
Pengfei Li, Yu~Hua, Pengfei Zuo, Zhangyu Chen, and Jiajie Sheng.
\newblock Rolex: A scalable rdma-oriented learned key-value store for
  disaggregated memory systems.
\newblock In {\em 21st USENIX Conference on File and Storage Technologies (FAST
  23)}, pages 99--114, 2023.

\bibitem{faiss}
A~library for efficient~similarity search and clustering of~dense vectors.
\newblock https://github.com/facebookresearch/faiss.

\bibitem{liu2024retrievalattention}
Di~Liu, Meng Chen, Baotong Lu, Huiqiang Jiang, Zhenhua Han, Qianxi Zhang,
  Qi~Chen, Chengruidong Zhang, Bailu Ding, Kai Zhang, et~al.
\newblock Retrievalattention: Accelerating long-context llm inference via
  vector retrieval.
\newblock {\em arXiv preprint arXiv:2409.10516}, 2024.

\bibitem{outback}
Yi~Liu, Minghao Xie, Shouqian Shi, Yuanchao Xu, Heiner Litz, and Chen Qian.
\newblock Outback: Fast and communication-efcient index for key-value store on
  disaggregated memory.
\newblock {\em {PVLDB}}, 18(2):335 -- 348, 2024.

\bibitem{lu2024dex}
Baotong Lu, Kaisong Huang, Chieh-Jan~Mike Liang, Tianzheng Wang, and Eric Lo.
\newblock Dex: Scalable range indexing on disaggregated memory.
\newblock {\em Proceedings of the VLDB Endowment}, 17(10):2603--2616, 2024.

\bibitem{luo2024chime}
Xuchuan Luo, Jiacheng Shen, Pengfei Zuo, Xin Wang, Michael~R Lyu, and Yangfan
  Zhou.
\newblock Chime: A cache-efficient and high-performance hybrid index on
  disaggregated memory.
\newblock In {\em Proceedings of the ACM SIGOPS 30th Symposium on Operating
  Systems Principles}, pages 110--126, 2024.

\bibitem{luo2023smart}
Xuchuan Luo, Pengfei Zuo, Jiacheng Shen, Jiazhen Gu, Xin Wang, Michael~R Lyu,
  and Yangfan Zhou.
\newblock Smart: A high-performance adaptive radix tree for disaggregated
  memory.
\newblock In {\em 17th USENIX Symposium on Operating Systems Design and
  Implementation (OSDI 23)}, pages 553--571, 2023.

\bibitem{hnsw}
Yu~A Malkov and Dmitry~A Yashunin.
\newblock Efficient and robust approximate nearest neighbor search using
  hierarchical navigable small world graphs.
\newblock {\em IEEE transactions on pattern analysis and machine intelligence},
  42(4):824--836, 2018.

\bibitem{roce}
Radhika Mittal, Alexander Shpiner, Aurojit Panda, Eitan Zahavi, Arvind
  Krishnamurthy, Sylvia Ratnasamy, and Scott Shenker.
\newblock Revisiting network support for rdma.
\newblock In {\em Proceedings of the 2018 Conference of the ACM Special
  Interest Group on Data Communication}, pages 313--326, 2018.

\bibitem{pan2024survey}
James~Jie Pan, Jianguo Wang, and Guoliang Li.
\newblock Survey of vector database management systems.
\newblock {\em The VLDB Journal}, 33(5):1591--1615, 2024.

\bibitem{mooncake}
Ruoyu Qin, Zheming Li, Weiran He, Mingxing Zhang, Yongwei Wu, Weimin Zheng, and
  Xinran Xu.
\newblock Mooncake: A kvcache-centric disaggregated architecture for llm
  serving.
\newblock {\em arXiv preprint arXiv:2407.00079}, 2024.

\bibitem{kdtree}
Parikshit Ram and Kaushik Sinha.
\newblock Revisiting kd-tree for nearest neighbor search.
\newblock In {\em Proceedings of the 25th acm sigkdd international conference
  on knowledge discovery \& data mining}, pages 1378--1388, 2019.

\bibitem{hm-ann}
Jie Ren, Minjia Zhang, and Dong Li.
\newblock Hm-ann: Efficient billion-point nearest neighbor search on
  heterogeneous memory.
\newblock In H.~Larochelle, M.~Ranzato, R.~Hadsell, M.F. Balcan, and H.~Lin,
  editors, {\em Advances in Neural Information Processing Systems}, volume~33,
  pages 10672--10684. Curran Associates, Inc., 2020.

\bibitem{shan2018legoos}
Yizhou Shan, Yutong Huang, Yilun Chen, and Yiying Zhang.
\newblock Legoos: A disseminated, distributed os for hardware resource
  disaggregation.
\newblock In {\em 13th USENIX Symposium on Operating Systems Design and
  Implementation (OSDI 18)}, pages 69--87, 2018.

\bibitem{shen2023ditto}
Jiacheng Shen, Pengfei Zuo, Xuchuan Luo, Yuxin Su, Jiazhen Gu, Hao Feng,
  Yangfan Zhou, and Michael~R Lyu.
\newblock Ditto: An elastic and adaptive memory-disaggregated caching system.
\newblock In {\em Proceedings of the 29th Symposium on Operating Systems
  Principles}, pages 675--691, 2023.

\bibitem{shen2023fusee}
Jiacheng Shen, Pengfei Zuo, Xuchuan Luo, Tianyi Yang, Yuxin Su, Yangfan Zhou,
  and Michael~R Lyu.
\newblock Fusee: A fully memory-disaggregatedkey-value store.
\newblock In {\em 21st USENIX Conference on File and Storage Technologies (FAST
  23)}, pages 81--98, 2023.

\bibitem{wentao}
Wentao Shi, Xiangnan He, Yang Zhang, Chongming Gao, Xinyue Li, Jizhi Zhang,
  Qifan Wang, and Fuli Feng.
\newblock Large language models are learnable planners for long-term
  recommendation.
\newblock In {\em Proceedings of the 47th International ACM SIGIR Conference on
  Research and Development in Information Retrieval}, pages 1893--1903, 2024.

\bibitem{su2024vexless}
Yongye Su, Yinqi Sun, Minjia Zhang, and Jianguo Wang.
\newblock Vexless: A serverless vector data management system using cloud
  functions.
\newblock {\em Proceedings of the ACM on Management of Data}, 2(3):1--26, 2024.

\bibitem{tsai2020disaggregating}
Shin-Yeh Tsai, Yizhou Shan, and Yiying Zhang.
\newblock Disaggregating persistent memory and controlling them remotely: An
  exploration of passive disaggregated key-value stores.
\newblock In {\em 2020 USENIX Annual Technical Conference (USENIX ATC 20)},
  pages 33--48, 2020.

\bibitem{wang2020semeru}
Chenxi Wang, Haoran Ma, Shi Liu, Yuanqi Li, Zhenyuan Ruan, Khanh Nguyen,
  Michael~D Bond, Ravi Netravali, Miryung Kim, and Guoqing~Harry Xu.
\newblock Semeru: A memory-disaggregated managed runtime.
\newblock In {\em 14th USENIX Symposium on Operating Systems Design and
  Implementation (OSDI 20)}, pages 261--280, 2020.

\bibitem{wang2024vector}
Jianguo Wang, Eric Hanson, Guoliang Li, Yannis Papakonstantinou, Harsha
  Simhadri, and Charles Xie.
\newblock Vector databases: What's really new and what's next?(vldb 2024
  panel).
\newblock {\em Proceedings of the VLDB Endowment}, 17(12):4505--4506, 2024.

\bibitem{wang2021milvus}
Jianguo Wang, Xiaomeng Yi, Rentong Guo, Hai Jin, Peng Xu, Shengjun Li, Xiangyu
  Wang, Xiangzhou Guo, Chengming Li, Xiaohai Xu, et~al.
\newblock Milvus: A purpose-built vector data management system.
\newblock In {\em Proceedings of the 2021 International Conference on
  Management of Data}, pages 2614--2627, 2021.

\bibitem{wang2023disaggregated}
Jianguo Wang and Qizhen Zhang.
\newblock Disaggregated database systems.
\newblock In {\em Companion of the 2023 International Conference on Management
  of Data}, pages 37--44, 2023.

\bibitem{wang2022sherman}
Qing Wang, Youyou Lu, and Jiwu Shu.
\newblock Sherman: A write-optimized distributed b+ tree index on disaggregated
  memory.
\newblock In {\em Proceedings of the 2022 international conference on
  management of data}, pages 1033--1048, 2022.

\bibitem{wang2022case}
Ruihong Wang, Jianguo Wang, Stratos Idreos, M~Tamer {\"O}zsu, and Walid~G Aref.
\newblock The case for distributed shared-memory databases with rdma-enabled
  memory disaggregation.
\newblock {\em arXiv preprint arXiv:2207.03027}, 2022.

\bibitem{wang2024rcmp}
Zhonghua Wang, Yixing Guo, Kai Lu, Jiguang Wan, Daohui Wang, Ting Yao, and
  Huatao Wu.
\newblock Rcmp: Reconstructing rdma-based memory disaggregation via cxl.
\newblock {\em ACM Transactions on Architecture and Code Optimization},
  21(1):1--26, 2024.

\bibitem{wei2023characterizing}
Xingda Wei, Rongxin Cheng, Yuhan Yang, Rong Chen, and Haibo Chen.
\newblock Characterizing off-path smartnic for accelerating distributed
  systems.
\newblock In {\em 17th USENIX Symposium on Operating Systems Design and
  Implementation (OSDI 23)}, pages 987--1004, 2023.

\bibitem{lighttraffic}
Yipeng Xing, Yongkun Li, Zhiqiang Wang, Yinlong Xu, and John~CS Lui.
\newblock Lighttraffic: On optimizing cpu-gpu data traffic for efficient
  large-scale random walks.
\newblock In {\em 2023 IEEE 39th International Conference on Data Engineering
  (ICDE)}, pages 882--895. IEEE, 2023.

\bibitem{zhang2024motor}
Ming Zhang, Yu~Hua, and Zhijun Yang.
\newblock Motor: Enabling multi-versioning for distributed transactions on
  disaggregated memory.
\newblock In {\em 18th USENIX Symposium on Operating Systems Design and
  Implementation (OSDI 24)}, pages 801--819, 2024.

\bibitem{zhang2022ford}
Ming Zhang, Yu~Hua, Pengfei Zuo, and Lurong Liu.
\newblock Ford: Fast one-sided rdma-based distributed transactions for
  disaggregated persistent memory.
\newblock In {\em 20th USENIX Conference on File and Storage Technologies (FAST
  22)}, pages 51--68, 2022.

\bibitem{zhang2019grip}
Minjia Zhang and Yuxiong He.
\newblock Grip: Multi-store capacity-optimized high-performance nearest
  neighbor search for vector search engine.
\newblock In {\em Proceedings of the 28th ACM International Conference on
  Information and Knowledge Management}, pages 1673--1682, 2019.

\bibitem{auncel}
Zili Zhang, Chao Jin, Linpeng Tang, Xuanzhe Liu, and Xin Jin.
\newblock Fast, approximate vector queries on very large unstructured datasets.
\newblock In {\em 20th USENIX Symposium on Networked Systems Design and
  Implementation (NSDI 23)}, pages 995--1011, 2023.

\bibitem{rummy}
Zili Zhang, Fangyue Liu, Gang Huang, Xuanzhe Liu, and Xin Jin.
\newblock Fast vector query processing for large datasets beyond gpu memory
  with reordered pipelining.
\newblock In {\em 21st USENIX Symposium on Networked Systems Design and
  Implementation (NSDI 24)}, pages 23--40, 2024.

\bibitem{hidpu}
Wenbin Zhu, Zhaoyan Shen, Qian Wei, Renhai Chen, Xin Yao, Dongxiao Yu, and Zili
  Shao.
\newblock Hidpu: A dpu-oriented hybrid indexing scheme for disaggregated
  storage systems.
\newblock In {\em 23rd USENIX Conference on File and Storage Technologies (FAST
  25)}, pages 271--285, 2025.

\bibitem{zuo2022race}
Pengfei Zuo, Qihui Zhou, Jiazhao Sun, Liu Yang, Shuangwu Zhang, Yu~Hua, James
  Cheng, Rongfeng He, and Huabing Yan.
\newblock Race: one-sided rdma-conscious extendible hashing.
\newblock {\em ACM Transactions on Storage (TOS)}, 18(2):1--29, 2022.

\end{thebibliography}

\end{document}